\documentclass{mn}
\usepackage{amssymb}

\usepackage{sw20mnra}

\newtheorem{theorem}{Theorem}
\newtheorem{acknowledgement}[theorem]{Acknowledgement}

\input{tcilatex}

\begin{document}

\title{Present limits to cosmic bubbles from the COBE-DMR three point correlation
function.}
\author{P. S. Corasaniti$^{2,1}$, L. Amendola$^{1}$, F. Occhionero$^{1}$ \\
%EndAName
$^{1}$Osservatorio Astronomico di Roma, Viale del Parco Mellini 84, 00136
Roma, Italy\\
$^{2}$Centre for Theoretical Physics, CPES, University of Sussex, Brighton BN1 9QJ,
United Kingdom}
\maketitle

\begin{abstract}
The existence of large scale voids in several galaxy surveys suggests the
occurrence of an inflationary first order phase transition. This process
generates primordial bubbles that, before evolving into the present voids,
leave at decoupling a non--Gaussian imprint on the CMB.

In this paper we evaluate an analytical expression of the collapsed three
point correlation function from the bubble temperature fluctuations.
Comparing the results with COBE-DMR measures, we obtain upper limits on the
allowed non-Gaussianity and hence on the bubble parameters.
\end{abstract}

\shortauthor{P.S. Corasaniti, L. Amendola \& F. Occhionero}
\keyphrases{\textit{(Cosmology:)} Cosmic microwave background}

\section{Introduction}

In the recent past the number of papers devoted to non-Gaussian anisotropies
on the Cosmic Microwave Background (CMB) has increased dramatically. This
new investigation field is, in fact, a powerful tool to distinguish between
the theories of structure formation based on inflation and those based on
topological defects. Quantum fluctuations produced in inflationary models
are scale invariant and have a Gaussian distribution. Thus we expect that
three-point correlation function of the CMB temperature vanish (Falk et al.
1993; Luo \& Schramm 1993; Gangui \& al. 1994).

On the contrary in models with topological defects the primordial density
perturbations are scale dependent and non-Gaussian (Avelino et al. 1998):
hence we expect some deviations from Gaussianity in higher order correlation
functions. In this context we may also include the extended inflation model
(La \& Steinhardt 1989), because during the inflationary epoch we have a
first order phase transition, that generates bubbles of true vacuum. These
voids contribute together with ordinary quantum fluctuations to structure
formation. This possibility has been investigated\ (Occhionero \&\ Amendola
1994; Amendola et al. 1996): it has been shown (Occhionero et al. 1984,
1997) that primordial bubbles may be associated with the observation of
large scale voids in several galaxy surveys (Kirshner et al. 1981; de
Lapparent et al. 1989; da Costa et al. 1996; El Ad, Piran \& da Costa 1996,
1997). Since these defects can also produce non-Gaussian anisotropies on the
CMB, we may obtain some limits on the bubble parameters comparing
observations with non-Gaussian predictions.

So far differents statistical tests have been applied to COBE-DMR sky maps
(Kogut et al. 1996) and the results have been in agreement with the Gaussian
models. Although recently two groups have detected non-Gaussian signal in
COBE data (Ferreira et al. 1998; Pando et al. 1998), subsequently this has
been shown to derive from a systematic effect in the data (Banday et al.
1999). On the other hand, Bromley \& Tegmark ({1999)} tried to argue that
the COBE 4 year data were in fact Gaussian.

In this paper we compare the level of non-Gaussianity produced in bubble
models with the COBE data. We evaluate the three point correlation function
in CDM models that contain also primordial bubbles. Comparing the numerical
results with the COBE-DMR measures (Kogut et al. 1996; see also Hinshaw et
al. 1994, 1995) we obtain upper limits on the parameters of the voids in
agreement with galaxy surveys observations.

\section{Method}

The imprints of bubbles on the CMB has been studied in several papers
(Baccigalupi, Amendola \&\ Occhionero 1997 ; Amendola, Baccigalupi \&
Occhionero 1998; Baccigalupi \& Perrotta 1999). The presence of the
primordial voids induces a Sachs-Wolfe effect and an acoustic perturbation
propagating up to the sound horizon on the photon distribution. As a
consequence, the induced temperature fluctuations of individual bubbles are
composed of a central spot and some concentric hotter isothermal rings; this
pattern has been calculated by numerical integration of the Boltzmann
equations in Amendola, Baccigalupi \& Occhionero (1998). It is found that
the bubble signal depends on the radius $R$ of the void and on the central
(negative) density contrast $\delta $. We shall distribute $N$ voids on the
CMB sky and use the fraction $X$ of the space that the voids fill today as a
free parameter, where $X=NR^{3}/3L_{h}^{2}\Delta L_{h}$ (Amendola et al.
1998) where $L_{h}$ is the horizon radius and $\Delta L_{h}$ is the
thickness of the last scattering surface. We consider bubbles of size $%
R=30h^{-1}$Mpc at decoupling: due to their overcoming growth, these voids
have today radii around $20\sim 60h^{-1}$ Mpc, like those observed in galaxy
surveys (da Costa et al. 1996; El Ad et al. 1996; 1997).

In simulated COBE maps, due to the low resolution of the satellite, the
signal of the individual bubbles looks like dark spots confused amidst
Gaussian anisotropies; their effect appears only when caculating the
correlation functions of maps containing many bubbles. The temperature
fluctuation may be decoupled in two terms: 
\begin{equation}
\Delta \left( \theta ,\varphi \right) =\Delta _{Gauss}\left( \theta ,\varphi
\right) +\Delta _{V}\left( \theta ,\varphi \right) .
\end{equation}
The first term $\Delta _{Gauss}\left( \theta ,\varphi \right) $ is the\
Gaussian temperature fluctuation field produced by the primary anisotropies;
the second term is the voids signal, that vanishes in directions where there
are not bubbles. In order to compare the predictions of the model with
experimental data, we calculate the collapsed three-point function 
\begin{eqnarray}
C_{3}(\alpha ) &=&\frac{1}{4\pi }\sum_{l_{1},l_{2},l_{3}}%
\sum_{m_{1},m_{2},m_{3}}P_{l_{1}}(\cos \alpha
)a_{l_{1}}^{m_{1}}a_{l_{2}}^{m_{2}}a_{l_{3}}^{m_{3}}  \nonumber \\
&&\times \mathcal{W}_{l_{1}}\mathcal{W}_{l_{2}}\mathcal{W}_{l_{3}}\mathcal{H}%
_{l_{1}l_{2}l_{3}}^{m_{1}m_{2}m_{3}}.  \label{collapsed}
\end{eqnarray}
where $\mathcal{W}^{l}$ is the window function of the experiment, $P_{l}$
are the Legendre polynomials, $a_{l}^{m}$ are the multipole coefficients of
the spherical harmonic expansion and where

\begin{eqnarray}
\mathcal{H}_{l_{1}l_{2}l_{3}}^{m_{1}m_{2}m_{3}} &=&(-1)^{m_{1}}\frac{\sqrt{%
(2l_{1}+1)(2l_{2}+1)(2l_{3}+1)}}{\sqrt{4\pi }}\times  \nonumber \\
&&\times \binom{l_{1}\text{ }l_{2}\text{ }l_{3}}{0\text{ }0\text{ }0}\binom{%
l_{1\text{ }}l_{2}\text{ }l_{3}}{m_{1}m_{2}m_{3}}.
\end{eqnarray}
with $\binom{l_{1\text{ }}l_{2}\text{ }l_{3}}{m_{1}m_{2}m_{3}}$ the Wigner
3J symbol. We compare $C_{3}(\alpha )$ to the pseudo three-point collapsed
function of Kogut et al. (1996): obviously the two are identical due to the
absence of noise in our case. Since the Gaussian term and the signal of the
bubbles are not correlated, we may write the three-point correlation
function as sum of two separate contributions: 
\begin{equation}
C_{3}=C_{3}^{Gauss}+C_{3}^{V}.
\end{equation}
The contribution to $C_{3}$ from gaussian fluctuations is not zero. This
contribution may arise from non-linearities in the inflationary dynamics or
from non-linear growth of the perturbations. However, using the analytical
expression for the $C_{3}^{Gauss}(\alpha )$ computed in Gangui et al. (1994)
it can be seen that the level of non-Gaussianity produced by the
non-linearities in the inflation dynamics is smaller than that arising from
the non-linear growth (Mollerach et al. 1995), and that the latter is much
smaller that produced by the bubbles. In fact, on the angular scales probed
by COBE-DMR, Mollerach et al. (1995) found an amplitude $\left\langle
C_{3}^{R-S}(\alpha )\right\rangle \sim 0.1$ $\mu K^{3}$, while we find that
the contribution of the voids is larger by several orders of magnitude: $%
\left\langle C_{3}^{V}(\alpha )\right\rangle \sim 10^{4}$ $\mu K^{3}$.
Therefore, we neglect $C_{3}^{Gauss}$ in the following.

To calculate $C_{3}^{V}(\alpha )$ we use the same approch of texture-spot
anisotropies (Magueijo 1995, Gangui \& Mollerach 1996, 1997). The
temperature fluctuations produced by a random distribution of bubbles, in
the $\hat{\gamma}$ direction, is simply the superposition of the signal of
all the bubbles, and can be written as $\Delta _{V}(\hat{\gamma}%
)=\sum_{n}b_{n}f_{n}(\hat{\gamma}_{n},\alpha )$. Here, the signal of the
n-th bubble has been decomposed as a overall amplitude $b_{n}$
(corresponding to the central temperature fluctuation) and a density profile 
$f_{n}(\hat{\gamma}_{n},\alpha )$ \ where $\alpha $ is the angle measured
from the bubble center. The dependence on the parameters $R_{n}$ and $\delta
_{n}$ is contained only in the amplitude; it is to be expected that this
dependence is linear in $\delta _{n}$ and quadratic in $R_{n}$, since the
central temperature fluctuation is dominated by the Sachs-Wolfe effect. The
expression $b_{n}=\delta _{n}(R_{n}/20H^{-1})^{2}$ is indeed an accurate fit
in the range we are interested (see Amendola et al. 1998). The profile $%
f_{n}(\hat{\gamma}_{n},\alpha )$ contains the full effect of the acoustic
oscillations and the adiabatic fluctuations, and has been obtained
numerically in Baccigalupi \& Perrotta (1999).

We expand $\Delta _{V}$ in spherical harmonics and obtain the multipole
coefficients 
\begin{equation}
a_{l}^{m}=\frac{4\pi }{2l+1}\sum_{n}b_{n}F_{n}^{l}Y_{l}^{m^{\ast }}(\hat{%
\gamma}_{n}),  \label{alm}
\end{equation}
where $F_{n}^{l}$ is the Legendre trasform of the intensity profile, 
\begin{equation}
F_{n}^{l}=\frac{2l+1}{2}\int d\Omega _{\alpha }f_{n}(\hat{\gamma}_{n},\alpha
)P_{l}(\cos \alpha ).
\end{equation}
Inserting (5) in (2) the collapsed function reduces
to: 
\begin{eqnarray}
C_{3}^{V}(\alpha ) &=&4\pi \sum_{l_{1},l_{2},l_{3}}P_{l_{1}}(\cos \alpha )%
\mathcal{W}^{l_{1}}\mathcal{W}^{l_{2}}\mathcal{W}^{l_{3}}\mathbf{J}%
^{l_{1}l_{2}l_{3}}  \nonumber \\
&&\times
\sum_{n_{1},n_{2},n_{3}}b_{n_{1}}b_{n_{2}}b_{n_{3}}F_{n_{1}}^{l_{1}}F_{n_{2}}^{l_{2}}F_{n_{3}}^{l_{3}},
\label{c3}
\end{eqnarray}
where $\mathbf{J}^{l_{1}l_{2}l_{3}}$ represents 
\begin{equation}
\mathbf{J}^{l_{1}l_{2}l_{3}}=\binom{l_{1}l_{2}l_{3}}{000}^{2}
\end{equation}
We take the window function of COBE to be $\sim e^{-l(l+1)\sigma ^{2}/2}$,
with $\sigma =3.2%
%TCIMACRO{\UNICODE[m]{0xb0}}%
%BeginExpansion
{{}^\circ}%
%EndExpansion
$. We assume now that there are $N$ identical voids on the CMB sky.
Developing the sum on $n_{1}$, $n_{2}$ and $n_{3}$ we obtain three terms
that represent the contribution to the $C_{3}^{V}(\alpha ),$ when the bubble
signals are not correlated and when are correlated two by two or three by
three and etc. We take into account the correlation at lowest order, in
other words we consider just \ the first two terms. Then the mean value of (7) for a Poissonian bubble distribution on the sky is obtained
substituting the sum on the bubble index with an integral over the whole
sky. In fact, the number of bubbles in a circular ring centered on a single
bubble is proportional to angular extension of the ring, therefore we have: 
\begin{eqnarray}
\left\langle C_{3}^{V}(\alpha )\right\rangle  &=&4\pi \delta ^{3}\left( 
\frac{R}{20H^{-1}}\right) ^{6}N\sum_{l_{1},l_{2},l_{3}}P_{l_{1}}(\cos \alpha
)  \nonumber \\
&&\times \mathcal{W}^{l_{1}}\mathcal{W}^{l_{2}}\mathcal{W}^{l_{3}}\mathbf{J}%
^{l_{1}l_{2}l_{3}}\mathit{I}^{l_{1}l_{2}l_{3}},  \label{bol}
\end{eqnarray}
where 
\begin{equation}
\mathit{I}^{l_{1}l_{2}l_{3}}=F^{l_{1}}F^{l_{2}}F^{l_{3}}+\frac{3}{2}%
F^{l_{1}}F^{l_{2}}\int F^{l_{3}}(\theta )d(\cos \theta ),
\end{equation}
and 
\begin{equation}
F^{l_{3}}(\theta )=\int f(\theta +\alpha )P_{l_{3}}(\cos \alpha )d(\cos
\alpha ).
\end{equation}
\ \  Using the same approch, after a tedious calculation, we have found an
analitycal expression for the variance $\sigma _{V}^{2}(\alpha
)=\left\langle C_{3}^{V}(\alpha )^{2}\right\rangle -\left\langle
C_{3}^{V}(\alpha )\right\rangle ^{2}$, that we do not report for shortness.
We compare the experimental data with the behaviour of the $\left\langle
C_{3}^{V}(\alpha )\right\rangle $ for differents values of the parameters $%
\delta $ and $X$. When $\left\langle C_{3}^{V}(\alpha )\right\rangle \pm
\sigma _{V}(\alpha )$ is larger than COBE data plus the noise and cosmic
variance, we have some constraints on the parameters of our model.

\section{Results}

The COBE data has been taken from Kogut et al. (1996). We assume a fraction
of bubbles corresponding to $0.31<X<0.54$, consistent with da Costa et al.
(1997). We have computed the $C_{3}^{V}(\alpha )$ for $0.001<\delta <0.0026$%
, without dipole and quadrupole contribution, $l_{\min }=4$. In the figures
we report the behaviour of the $C_{3}^{V}(\alpha )$ for two values of $%
\delta =0.002,0.0012$. The oscillating behaviour of the plots is due to the
sum of the Legendre polynomials in (9). In the plots the errorbars
are the $\sigma (\alpha )$'s. The level of the cosmic variance $\sigma
(\alpha )$ generated from the model is very high for $\alpha <40%
%TCIMACRO{\UNICODE[m]{0xb0}}%
%BeginExpansion
{{}^\circ}%
%EndExpansion
$, while it is small on the large angular scales, $\alpha >45%
%TCIMACRO{\UNICODE[m]{0xb0}}%
%BeginExpansion
{{}^\circ}%
%EndExpansion
$, where the contribution of the lowest multipoles is small. In figure (1)
we have the model with $\delta =0.002$: we may note that for $X=0.54$ the
signal is larger than cosmic variance and the observed data points, while $%
X=0.31$, the plot is marginally consistent with the experimental data.\FRAME{%
ftbpFU}{3.1012in}{3.1116in}{0pt}{\Qcb{The points are the COBE data while the
thick lines are the cosmic variance of a Gaussian random field (Kogut et
al., 1996). The plots are models with $\protect\delta =0.002$ and $X=0.54$
(dashed line) and $X=0.31$ (thin line). The errorbars are the variance of
our models.}}{}{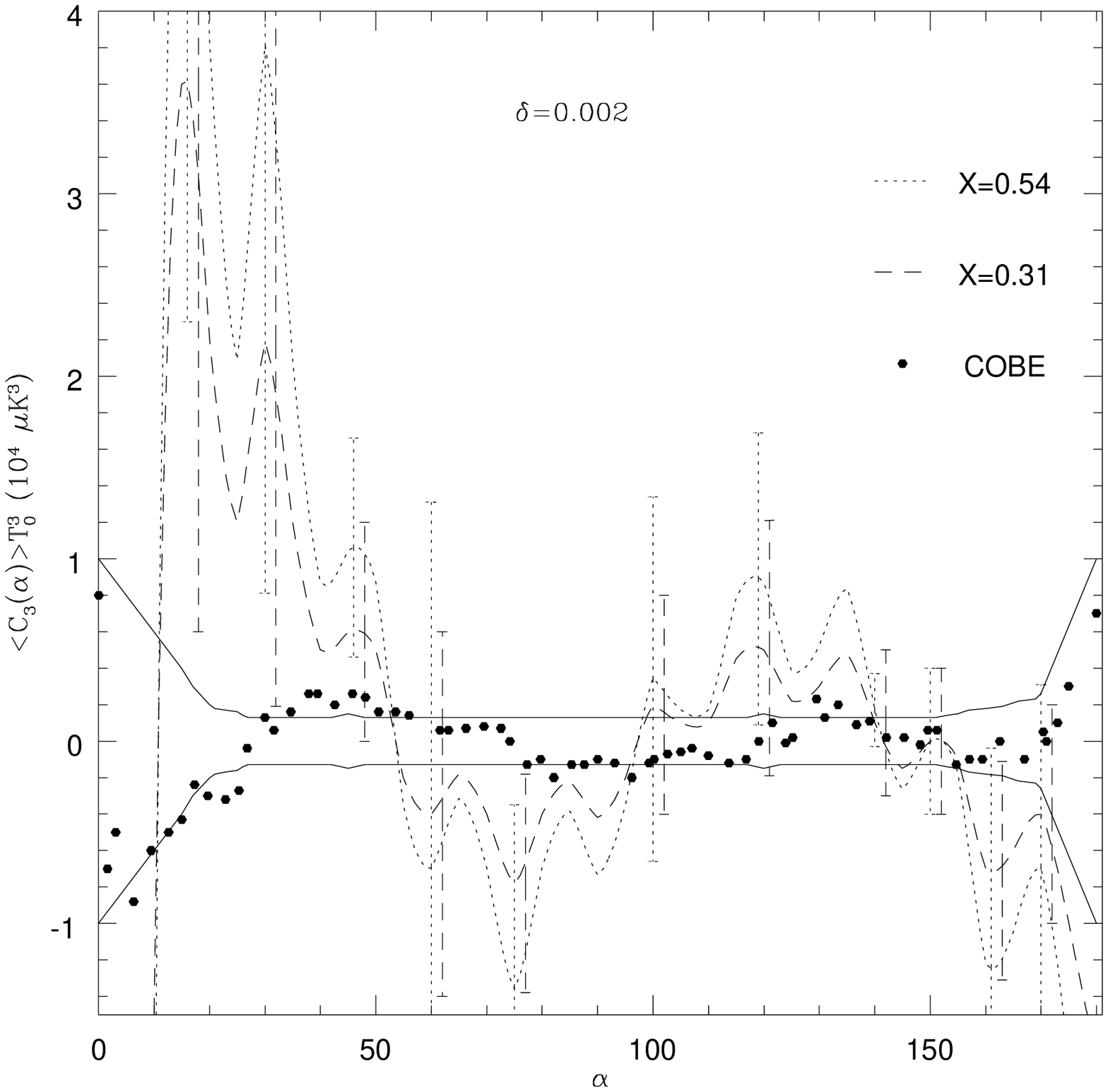}{\special{language "Scientific Word";type
"GRAPHIC";maintain-aspect-ratio TRUE;display "USEDEF";valid_file "F";width
3.1012in;height 3.1116in;depth 0pt;original-width 7.4884in;original-height
7.3838in;cropleft "0";croptop "1.0172";cropright "1";cropbottom "0";filename
'g002.ps';file-properties "XNPEU";}} In figure (2) we report the $\langle
C_{3}^{V}(\alpha )\rangle T_{0}^{3}$ for $\delta =0.0012$: it fits the COBE
data very well. \FRAME{ftbpFU}{3.1012in}{3.1116in}{0pt}{\Qcb{Same as in Fig.
1, but now $\protect\delta =0.0012$: agreement with observations is now
obtained.}}{}{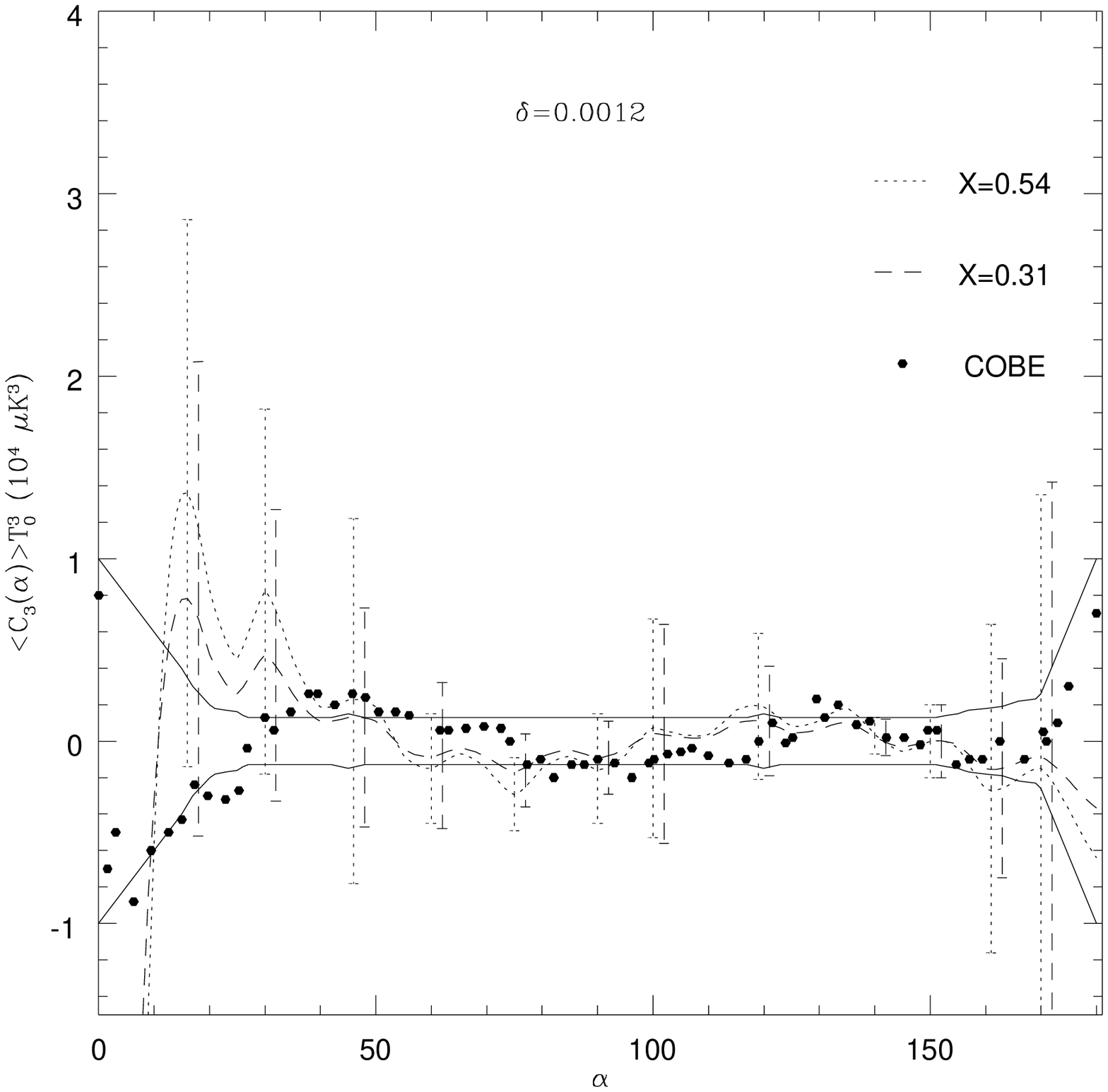}{\special{language "Scientific Word";type
"GRAPHIC";maintain-aspect-ratio TRUE;display "USEDEF";valid_file "F";width
3.1012in;height 3.1116in;depth 0pt;original-width 7.4884in;original-height
7.3838in;cropleft "0";croptop "1.0174";cropright "1";cropbottom "0";filename
'g0012.ps';file-properties "XNPEU";}} Notice that in the range $\alpha >50%
%TCIMACRO{\UNICODE[m]{0xb0}}%
%BeginExpansion
{{}^\circ}%
%EndExpansion
$ the behaviour of the collapsed function seems to follow the trend of the
COBE measures. Values of $\delta <0.0012$ produce a $C_{3}^{V}(\alpha )$
within the cosmic variance band and smaller than the COBE data. In this case
the observations do not impose constraints and we may obtain only an upper
limit on the value of $\delta $. We have applied a $\chi ^{2}$ analysis to
our models. In figure (3) we report the confidence regions with a confidence
level set to 99.9\% (grey region) and to 99.5\% (black region). \FRAME{ftbpFU%
}{3.0684in}{3.0684in}{0pt}{\Qcb{Confidence regions for different values of $%
\protect\delta $ and $X$, the confidence level is set to 99.9\% (grey
region) and to 99.5\% (black region). Models with $\protect\delta >0.0017$
are ruled out by the COBE measures.}}{}{regconfi.eps}{\special{language
"Scientific Word";type "GRAPHIC";maintain-aspect-ratio TRUE;display
"USEDEF";valid_file "F";width 3.0684in;height 3.0684in;depth
0pt;original-width 4.4789in;original-height 4.4789in;cropleft "0";croptop
"1";cropright "1";cropbottom "0";filename
'../../users/luca/work/texwork/stefano/Programmi/Wolfram
Research/regconfi.emf';file-properties "NPEU";}}We may note that all models
with $\delta \gtrsim 0.0017$ are ruled out by the experimental data. Then we
may conclude that although the bubbles produce a non-Gaussian signal on the
CMB, this is in agreement with the present observation provided that the
density contrast $\delta \leq 0.0017$ or $X\leq 54\%$. So we obtain a
constraint stronger than that found in Amendola et al. (1998), where the
bubble power spectrum was compared to the measures of the CAT experiment.
The next high resolution experiments, like MAP and Planck, and the recent
observations of Boomerang and Maxima should be able to detect the voids
signal on the CMB. In fact these missions can probe the multipoles $l>100$,
where the contribution of the bubbles is important, and the effects on $%
C_{3}(\alpha )$ may be large.

\section{Conclusion}

Several galaxy surveys found huge spherical voids in the matter
distribution, the galaxies lying in the surrounding shells: these structures
may be generated in inflationary models with first order phase transitions.
These bubbles produce a non-Gaussian signal on the CMB. We analyse this
signal developing an analytical expression for the three-point collapsed
function of a bubble distribution, using the formalism of Magueijo (1995).
Our free parameters are the density contrast and volume fraction of the
bubbles, while the radius $R$ is fixed to a value consistent with the galaxy
surveys. We compare the behaviour of the three point collapsed function for
the bubble model with the COBE data. We obtain a constraint on the value of $%
\delta $: in fact, the existence of the voids at decoupling is not in
contrast with the measures of the COBE three-point collapsed function,
provided $\delta \leq 0.0017$ or $X\leq 0.54$. This still leaves plenty of
room for the bubbles to cooperate efficiently to structure formation, both
via the central voids and via the possibility of shocking on the outher
shell: in fact a central density contrast of $0.001$ can still evolve
linearly in an empty void by today. More information will be obtained
comparing the results of the future high resolution experiments.

\begin{acknowledgement}
L.A. and F. O. acknowledge financial support from the Italian Ministry of
University Research and Scientific Technology
\end{acknowledgement}

\end{document}